\documentclass{aastex62}

\newcommand{\teff}{\mbox{T$_{\rm eff}$}}
\newcommand{\logg}{\mbox{log~{\it g}}}

\newcommand{\kmsec}{\mbox{km~s$^{\rm -1}$}}

\newcommand{\vsini}{\mbox{$v~{\rm sin}{\it i}$}}

\accepted{11/07/2018}

\shorttitle{Stellar rotation in the double MS of NGC\,1818: first
  spectroscopic analysis} 
\shortauthors{A.\,F. Marino, et al.} 

\begin{document}
\title{DIFFERENT STELLAR ROTATION IN THE TWO MAIN SEQUENCES OF THE YOUNG
  GLOBULAR CLUSTER NGC\,1818: FIRST DIRECT SPECTROSCOPIC
  EVIDENCE\footnote{Based on observations collected at the European
    Organisation for Astronomical Research in the Southern Hemisphere
    under ESO programme 0100.D-0520(A), and on observations with the
    NASA/ESA {\it Hubble Space Telescope}, obtained at the Space
    Telescope Science Institute, which is operated by AURA, Inc.,
    under NASA contract NAS 5-26555.}} 

\author{A.\ F.\,Marino} 
\affiliation{Research School of Astronomy \& Astrophysics, Australian National University, Canberra, ACT 2611, Australia} 
\author{N.\ Przybilla}
\affiliation{Institut f\"{u}r Astro- und Teilchenphysik, Universit\"{a}t Innsbruck, Technikerstrasse 25, 6020, Innsbruck, Austria} 
\author{A.\ P.\,Milone}
\affiliation{Dipartimento di Fisica e Astronomia ``Galileo Galilei'' - Univ. di Padova, Vicolo dell'Osservatorio 3, Padova, IT-35122}
\author{G.\ Da Costa}
\affiliation{Research School of Astronomy \& Astrophysics, Australian National University, Canberra, ACT 2611, Australia}  
\author{F.\ D'Antona}
\affiliation{INAF - Osservatorio Astronomico di Roma, Monte Porzio, Italy}
\author{A.\ Dotter}
\affiliation{Harvard-Smithsonian Center for Astrophysics, 60 Garden Street, Cambridge, MA 02138, USA}  
\author{A.\ Dupree}
\affiliation{Harvard-Smithsonian Center for Astrophysics, 60 Garden Street, Cambridge, MA 02138, USA}

\correspondingauthor{A.\ F.\,Marino}
\email{anna.marino@anu.edu.au}

\begin{abstract}  
We present a spectroscopic analysis of main sequence (MS) stars in the
young globular cluster NGC\,1818 (age$\sim$40~Myrs) in the Large
Magellanic Cloud.
Our photometric survey on Magellanic Clouds clusters has revealed that
NGC\,1818, similarly to the other young objects with
age$\lesssim$600~Myrs, displays not only an extended MS Turn-Off (eMSTO), as
observed in intermediate-age clusters (age$\sim$1-2~Gyrs), but also a split
MS. The most straightforward interpretation of the double MS is the
presence of two stellar populations: a sequence of slowly-rotating stars
lying on the blue-MS and a sequence of fast rotators, with rotation
close to the breaking speed, defining a red-MS.
We report the first direct spectroscopic measurements of projected rotational
velocities \vsini\ for the double MS, eMSTO and Be stars of a young cluster.
The analysis of line profiles includes non-LTE effects, required for
correctly deriving \vsini\ values. 
Our results suggest that:
{\it (i)} the mean rotation for blue- and red- MS stars is
\vsini=71$\pm$10~\kmsec\ ($\sigma$=37\kmsec) and  \vsini=202$\pm$23~\kmsec\
($\sigma$=91\kmsec), respectively; 
{\it (ii)} eMSTO stars have different \vsini, which are generally
lower than those inferred for red-MS stars, and 
{\it (iii)} as expected, Be stars display the highest \vsini\ values.
This analyis supports the idea that distinct rotational velocities
play an important role in the appearence of multiple stellar
populations in the color-magnitude diagrams of young clusters, and
poses new constraints to the current scenarios.
\end{abstract}


\section{Introduction}\label{sec:intro}

In the last decade, several papers based on high-precision photometry,
have revealed that the color-magnitude diagram (CMD) of most old
Galactic Globular Clusters (GGCs) is made of multiple sequences
(e.g. Milone et al.\,2017a). These
sequences have been observed along the entire CMD and correspond to
distinct stellar populations with different helium and C, N, O, Na
abundance (e.g. Marino et al.\,2008, 2016; Milone et al.\,2015a). 
To explain the observations, some authors have suggested
that GCs have experienced a prolonged star-formation history
(e.g.\,D'Antona et al.\,2016; Decressin et al.\,2007), 
while other work concludes that stars in GCs are
coeval and that the observed multiple populations are due to disk
accretion in pre main sequence (MS) stars (e.g. Bastian et al.\,2013). 
In general, the series of events that led from massive gas
clouds in the early Universe to the GCs we see today with their
multiple populations, remains a puzzle (Renzini et al.\,2015). 

Multiple stellar populations are not a peculiarity of old Milky Way
GCs. In the last decade high-accuracy photometry from the Hubble Space
Telescope ($HST$) has revealed that several 1-2~Gyr old star clusters in
both Magellanic Clouds (MCs) exhibit bimodal or extended main-sequence turn
offs (eMSTOs) and dual clumps (Mackey \& Broby Nielsen\,2007; Milone
et al.\,2009; Glatt et al.\,2008; Girardi et al.\,2009). In
particular, our $HST$ survey of star clusters in 
the MCs has shown that the eMSTO is a common feature of 
intermediate-age star clusters, as it has been observed in the majority
of the analyzed clusters (Milone et al.\,2009). Several
authors have thus suggested that eMSTO clusters are the counterparts
of the old GCs with multiple sequences (Conroy \& Spergel\,2011; 
Keller et al.\,2011) and that they provide a
laboratory to study multiple populations few hundreds-Myr after their
formation. 
However, while the MSTO of most intermediate-age star
clusters is widely broadened in color and magnitude, the MS seems
consistent with a single isochrone (Milone et al.\,2009). 

More recently, double MSs have been detected instead in very young clusters,
with ages between $\sim$50 and $\sim$600~Myrs (Milone et al.\,2015b,
2016, 2017b, 2018; Correnti et al.\,2017).
These young GCs do not only exhibit an eMSTO, but, differently from the
intermediate-age clusters, they also host a split MS, similar to that
observed in Milky Way globulars (e.g.\,Milone et al.\,2012; Milone\,2015). 

The most straightforward interpretation of the
eMSTO was an extended star-formation history with a duration of
100-500~Myr (e.g. Milone et al.\,2009; Goudfrooij et
al.\,2014). According to this scenario, young and
intermediate-age GCs have experienced a complex star-formation
history, with two or more stellar generations. This in turn would suggest
that young and intermediate-age MC clusters could be the counterpart of
old GGCs with their multiple stellar populations. As an
alternative, Bastian \& de Mink\,(2009) suggested that
stellar rotation can mimic an age spread and an eMSTO.

Comparison of the CMD of young clusters with isochrones suggests that
the split MS is best reproduced by two stellar populations with
distinct rotation: 
a non-rotating population of blue MS stars and a population of red MS
stars with rotational velocity close to the critical value (D'Antona et
al.\,2015; Milone et al.\,2016).  
One the most crucial topic in the field of multiple stellar populations is to
understand if the double MSs in young GCs are the counterpart of those
observed on old Milky Way globulars or are instead a different phenomenon.

Direct detection of rapidly rotating stars among the eMSTO was
provided by Dupree et al.\,(2017) from the analysis of high resolution
spectra of eMSTO stars in the young cluster NGC\,1866.
In this work we analyse spectroscopic data for stars in the GC
NGC\,1818 in the Large MC (LMC), in order to estimate the projected rotational
velocity \vsini. 
NGC\,1818 ($\rm {log}(M/M_{\odot})=4.41$\footnote{Taken from McLaughlin
\& van der Marel (2005) by assuming a Wilson (1975) profile.}), with
an age of $\sim$40~Myrs, is one of the very young GCs 
with a detected split MS (Milone et al.\,2018),
interpreted as composed by $\sim$30\% of non-rotating stars, on the
blue-MS (bMS) and $\sim$70\% of fast-rotators with rotation $\omega \sim 0.9
\omega_{\rm crit}$ on the red-MS (rMS), in the mag range $18.2\lesssim m_{\rm
  {F814W}} \lesssim 20.7$. An eMSTO plus a population of Be stars with
mag $15.5 \lesssim m_{\rm {F814W}} \lesssim 18.5$ have been also detected.
The layout of this paper is as follows: 
Sect.~\ref{sec:data} describes the photometric and spectroscopic data;
our results are presented in Sect.~\ref{sec:rotation} and discussed in
Sect.~\ref{sec:discussion}.

\section{Data}\label{sec:data}

\subsection{The photometric dataset\label{sec:phot_data}}

To investigate multiple sequences along the CMD of NGC\,1818 we used
the photometric and astrometric catalogue published by Milone et
al.\,(2018).
This catalogue includes {\it HST} photometry of NGC\,1818 stars in the
F336W, F656N, and F814W bands of the Ultraviolet and Visual Channel of
the Wide Field Camera 3. Flux and magnitude measurements have been
derived from the images collected as part of GO-13727 (PI J.\,Kalirai)
and GO-14710 (PI A.\,Milone) by using the methods of data analysis and
data reduction devoloped by Jay Anderson. Photometry has been
calibrated to the VEGA-mag system by using the photometric zero points
provided by the WFC3/UVIS web page (see Bedin et al.\,2005 for
details), while stellar positions are tranformed in the Gaia data
release 1 reference frame (Gaia collaboration et al.\,2016).
We refer the interested reader to the works by Milone et al.\,(2018),
Anderson et al.\,(2008) and references therein for details on the
database and on the data reduction.

\subsection{The spectroscopic dataset}\label{sec:spec_data}

The spectroscopic data consist of FLAMES/GIRAFFE data (Pasquini et
al.\,2002) collected under the ESO programme 0100.D-0520 (PI: Marino)
taken in Visitor mode during the nights 13/14/15 December 2017. 

The GIRAFFE fibres were used with the HR15N setup, covering the
spectral range from $\sim$647 to $\sim$679~nm with a resolving power of
$R=\lambda/\Delta\lambda\sim$19200. 
We have observed a total of 44 stars in the
2.7$\arcmin$$\times$2.7$\arcmin$ field of view 
around NGC\,1818 by using two FLAMES configurations.
The two configurations were observed in 15 and 13 different exposures
of $\sim$46~min.  
Data reduction involving bias subtraction, flat-field correction,
wavelength calibration, sky subtraction and continuum normalization 
has been done using the dedicated pipelines for GIRAFFE and {\sf
  IRAF}\footnote{IRAF is distributed by NOAO, which is operated by
  AURA, under cooperative agreement with the US National Science
  Foundation.} standard procedures. 
At the wavelength of the He\,{\sc i} 6678~\AA\, line, the typical
signal-to-noise ratio 
(S/N) of the fully reduced and combined spectra ranges from $\sim$30
for the faintest targets to $\sim$80 for the brightest ones. 

\begin{figure*}
   \centering
   \includegraphics[width=15cm]{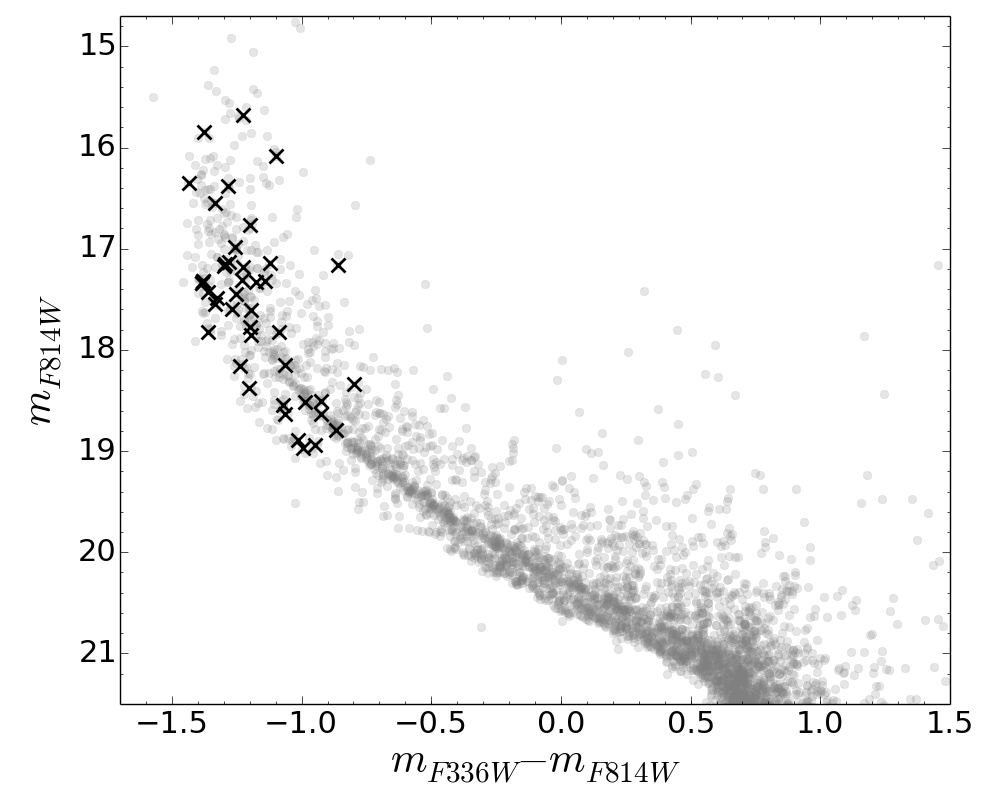}
      \caption{
$m_{\rm F814W}$-$(m_{\rm F336W}-m_{\rm F814W})$ CMD for NGC\,1818. Our
spectroscopic targets have been represented with black crosses.
       }
        \label{fig:targets}
   \end{figure*}

The observed stars, shown in Fig.~\ref{fig:targets}, have been
selected from the CMD described in 
Sect.~\ref{sec:phot_data}. Specifically, we have observed 
stars on both the bMS and
rMS, plus some objects in the eMSTO region, and 4 stars classified as Be (see
Sect.~\ref{sec:Be} for further details on their selection).
A few stars are observed with a redder color with respect to the rMS,
where binaries are expected.
In the following we discuss the observed stars on the basis of their
location on the CMD, plus spectroscopic information. Hence, bona-fide
bMS and rMS stars will be discussed separately from the eMSTO and
Be stars. Stars clearly showing double-line features in their spectra
are classified as binaries.
It is noteworthy to clarify that,
although our high-precision photometry allowed us to properly select
stars on different regions of the CMD with small errors, a residual
contamination is expected. In particular, from the CMD in
Fig.~\ref{fig:targets}, it is clear that there could be some overlap
between the rMS, binaries and/or Be stars. 
To avoid confusion, we keep this selection thorough the paper, and discuss
in each section the presence of possible contaminants.

The stellar spectral lines covered with the HR15N setup in the sample stars
are H$\alpha$, He\,{\sc i} $\lambda$6678\,{\AA} and in some cases the
C\,{\sc ii} 
doublet $\lambda\lambda$\,6578/82\,{\AA}. In addition,
spatially variable sharp nebular emission lines are visible, comprising 
H$\alpha$, [N\,{\sc ii}] $\lambda\lambda$\,6548/84\,{\AA} and [S\,{\sc ii}]
$\lambda\lambda$\,6717/31\,{\AA}, in particular in the fainter objects.

Heliocentric radial velocities $v_\mathrm{r,\odot}$ of the sample
stars were determined by 
cross-correlation of the observed spectra with appropriate synthetic
spectra (see below). 
Overall, the precision of the $v_\mathrm{r,\odot}$-determination for
the individual star is influenced by the number of useful lines
(H$\alpha$ and/or He\,{\sc i} $\lambda$6678\,{\AA} were considered),
the rotational broadening (sharper lines giving more precise results) and
the presence and location of the nebular emission features. After
application of a 2$\sigma$-clipping 
criterion with iteration depth 2, 33 out of 39 stars with $v_\mathrm{r,\odot}$
data remain for the determination of the cluster radial velocity of
311.1\,$\pm$\,3.9\,km\,s$^{-1}$ (1$\sigma$-scatter). 
The outliers, stars \#27, \#28, \#96, \#57, and \#77, 
which belong to our rMS and bMS selection, are assumed
being single-lined spectroscopic binaries 
(SB1) candidates. However, due to their similarly high
$v_\mathrm{r,\odot}$, while advising about their possible binarity,
we will treat them similarly to the other rMS and bMS samples.
Individual $v_\mathrm{r,\odot}$ corrected to the heliocentric system,
togheter with coordinates, basic $HST$ photometry for the 
observed NGC\,1818 stars are listed in Tab.~\ref{tab:data}.

\section{Rotation}\label{sec:rotation}

Projected rotational velocities \vsini\ of the sample stars were
determined by fitting the H$\alpha$ Doppler core and/or the He\,{\sc i}
$\lambda$6678\,{\AA} line. For this, a grid of hybrid
non local thermodynamic equilibrium (non-LTE) model spectra was
computed using the approach discussed by 
Nieva \& Przybilla (2007) and Przybilla et al. (2011).
In brief, model atmospheres in the relevant atmospheric parameter range 
as deduced from the CMD were computed using the {\sc Atlas9} code
(Kurucz 1993), which assumes plane-parallel geometry, chemical homogeneity,  
hydrostatic, radiative equilibria and LTE. 
The grid covers the effective temperature range
12\,000\,$\le$\,\teff\,$\le$\,23\,000\,K, surface gravities 
3.5\,$\le$\,\logg\,$\le$\,4.5 and helium abundances
0.07\,$\le$\,$y$\,$\le$0.10 (by number) at half solar metallicity, as
appropriate for the young stellar population of the LMC (e.g. Korn et al. 2002,
including one NGC\,1818 B-type star). 
We account for limb darkening 
by using linear limb-darkening coefficients taken from Wade \& Rucinski (1985).
Non-LTE level populations were
computed using {\sc Detail} and synthetic non-LTE and LTE spectra using
{\sc Surface} (Giddings~1981; Butler \& Giddings 1985, both  updated
and extended by K. Butler). We employed model atoms for hydrogen and helium as
described by Przybilla \& Butler (2004) and Przybilla (2005).

\begin{figure*}
   \centering
   \includegraphics[width=15cm]{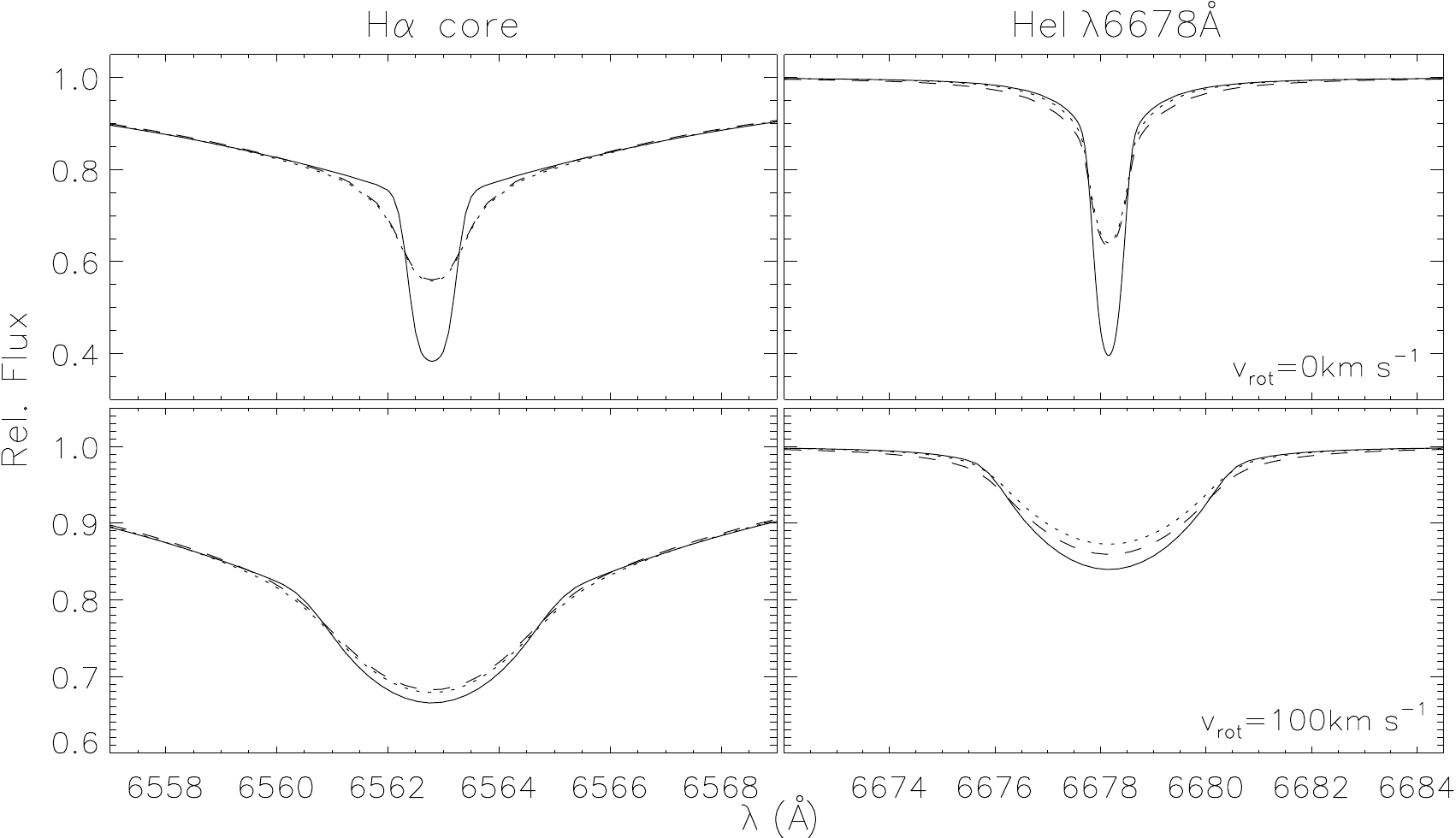}
      \caption{{\it Upper panels}: Comparison of non-LTE model
        profiles without rotation (full lines) with LTE profiles
        (dotted lines) for the same atmospheric parameters
($T_\mathrm{eff}$=\,22\,000\,K, $\log g$\,=\,4.00 and $y$\,=\,0.08). 
A second LTE profile for $y$\,=\,0.12 (dashed line) 
reproduces the equivalent width of the non-LTE He\,{\sc i} model profile.
{\it Lower panels}: the same after convolution with a rotational profile for
100\,km\,s$^{-1}$.
       }
        \label{fig:nlte}
   \end{figure*}

Both the deep and narrow H$\alpha$ Doppler core as well as the deep
He\,{\sc i} $\lambda$6678\,{\AA} line result from non-LTE effects
and cannot be reproduced by LTE calculations. The upper panels of 
Fig.~\ref{fig:nlte} show a comparison of profiles for the two lines for zero
rotation, in non-LTE and LTE, and also an LTE profile for
increased helium abundance that reproduces the equivalent width of 
the non-LTE He\,{\sc i} profile. The LTE line profiles are shallower and 
broader than the non-LTE profiles and the H$\alpha$ Doppler core is
largely insensitive to helium abundance changes (within reasonable
limits, here 50\%). The lower panels of Fig.~\ref{fig:nlte} show the
situation after convolution with a rotational profile for 100\,\kmsec. 
The different line profiles remain qualitatively different. An
LTE analysis can not derive a consistent \vsini\ values from both
lines simultaneously, and the values will differ from those obtained
by non-LTE fits. We conclude that the consideration of non-LTE effects 
is essential for deriving accurate and precise \vsini\ values from 
these two lines.  

We employed {\sc Spas} (Spectrum Plotting and Analysing Suite, Hirsch 2009)
for fitting model spectra to the observations. {\sc Spas}
provides the means to interpolate between model grid points for up
to three parameters simultaneously (here \teff, \logg\ and $y$) and
allows instrumental, rotational, and  (radial-tangential)   
macrobroadening functions to be applied to the resulting theoretical profiles. 
The program uses the downhill simplex algorithm  (Nelder \& Mead  1965)
to minimise $\chi^2$ in order to find a good fit to the observed spectrum. 

\begin{figure*}
   \centering
   \includegraphics[width=5.94cm]{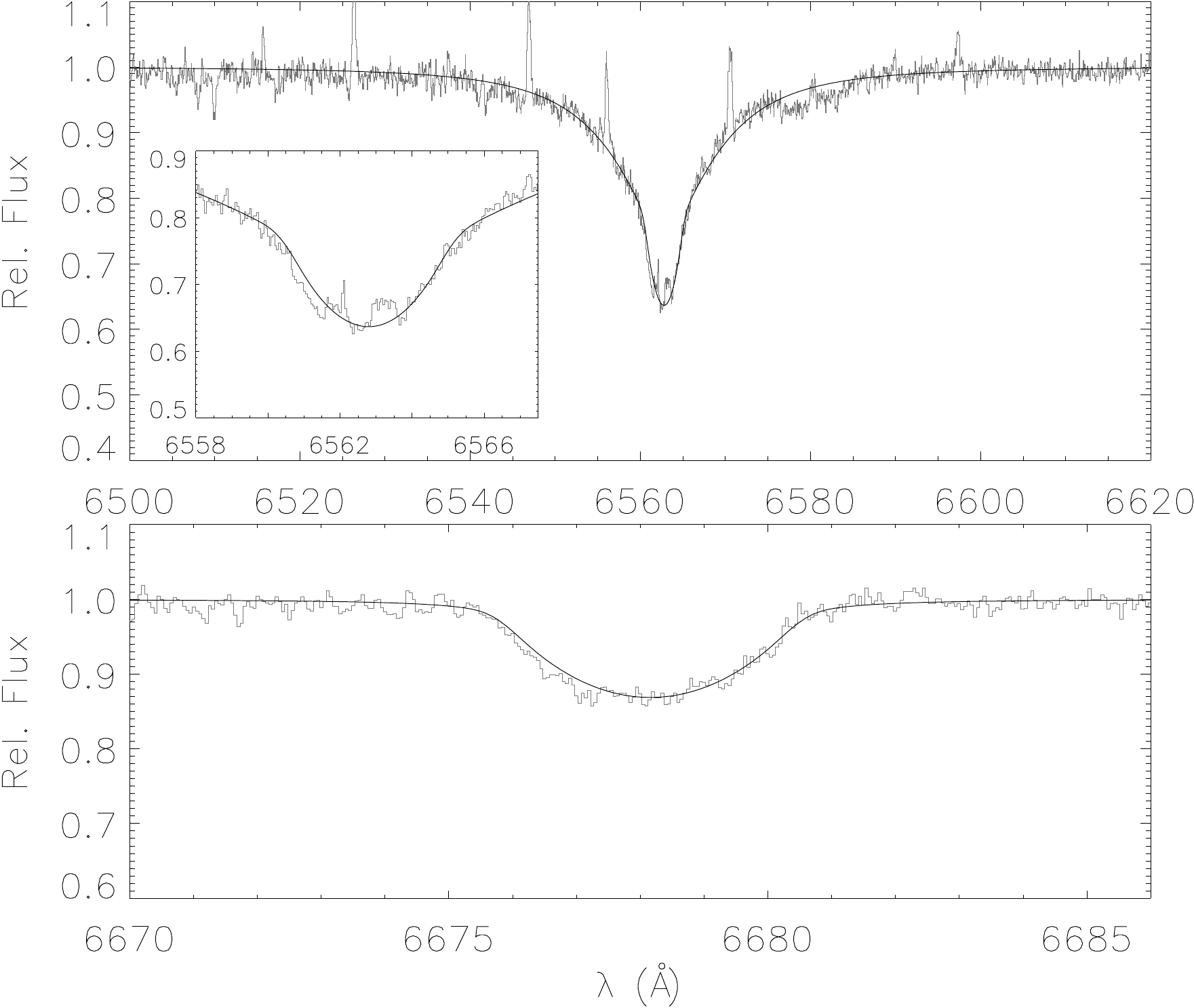} 
   \includegraphics[width=5.94cm]{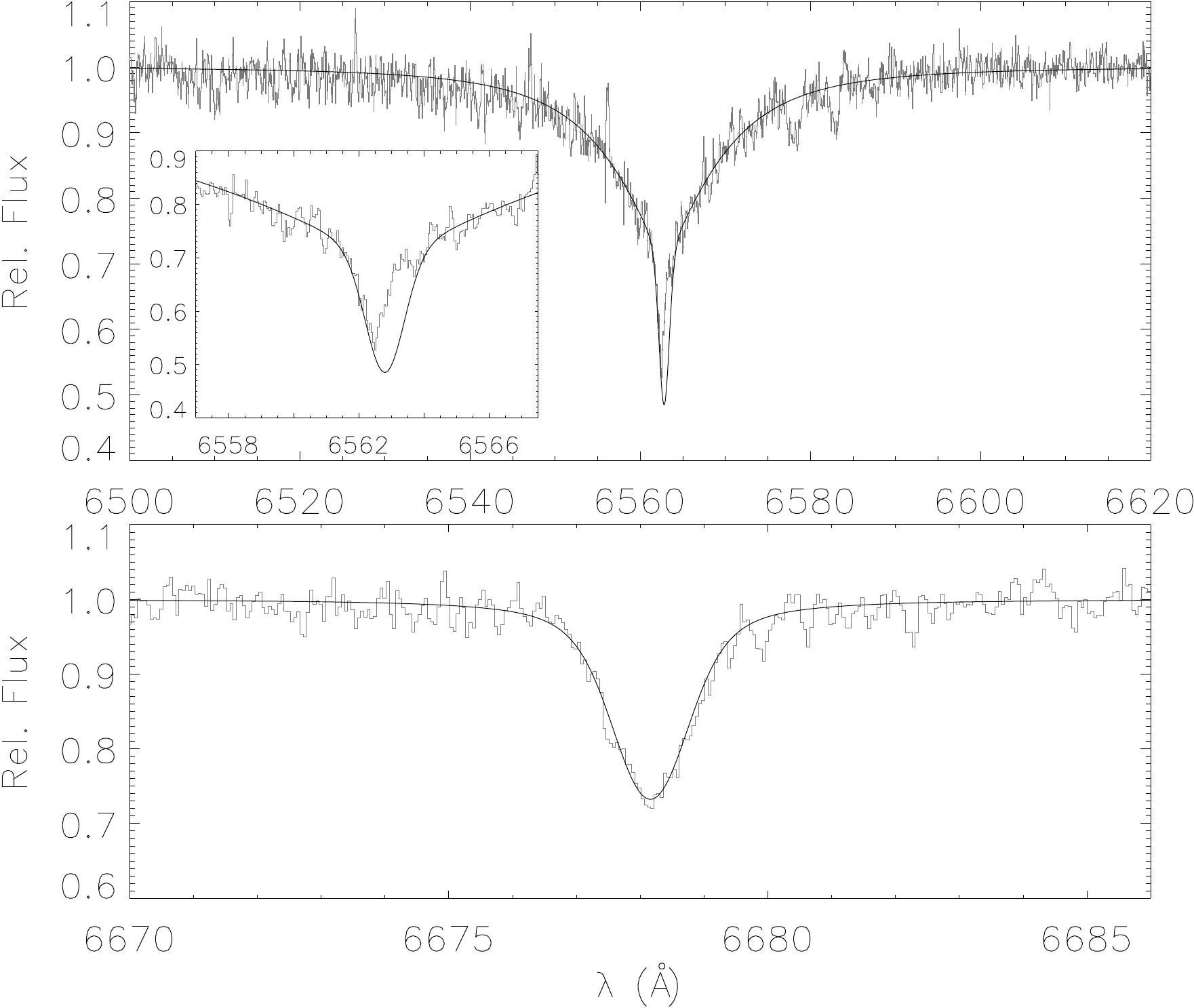}  
   \includegraphics[width=5.94cm]{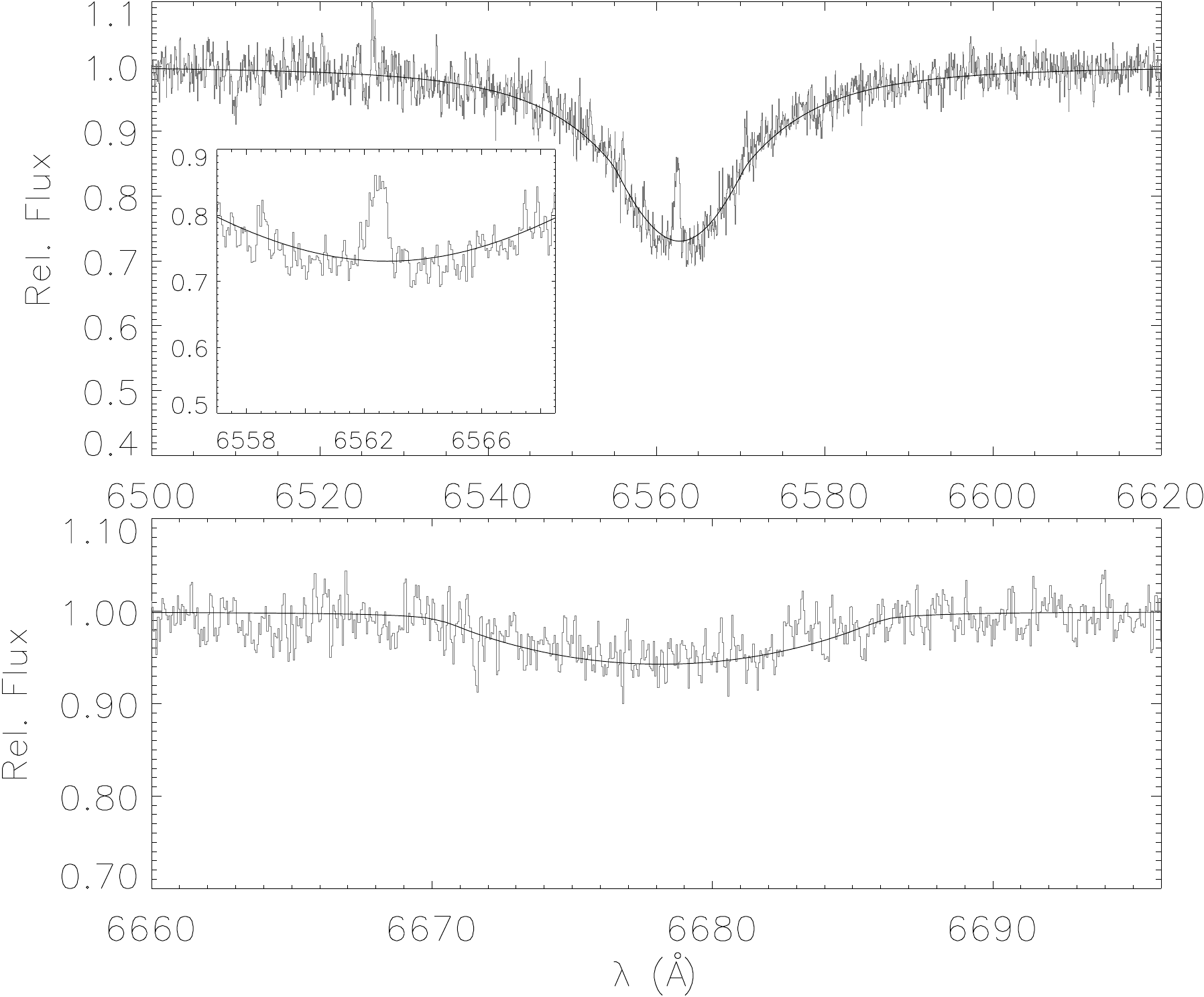} 
   \includegraphics[width=5.94cm]{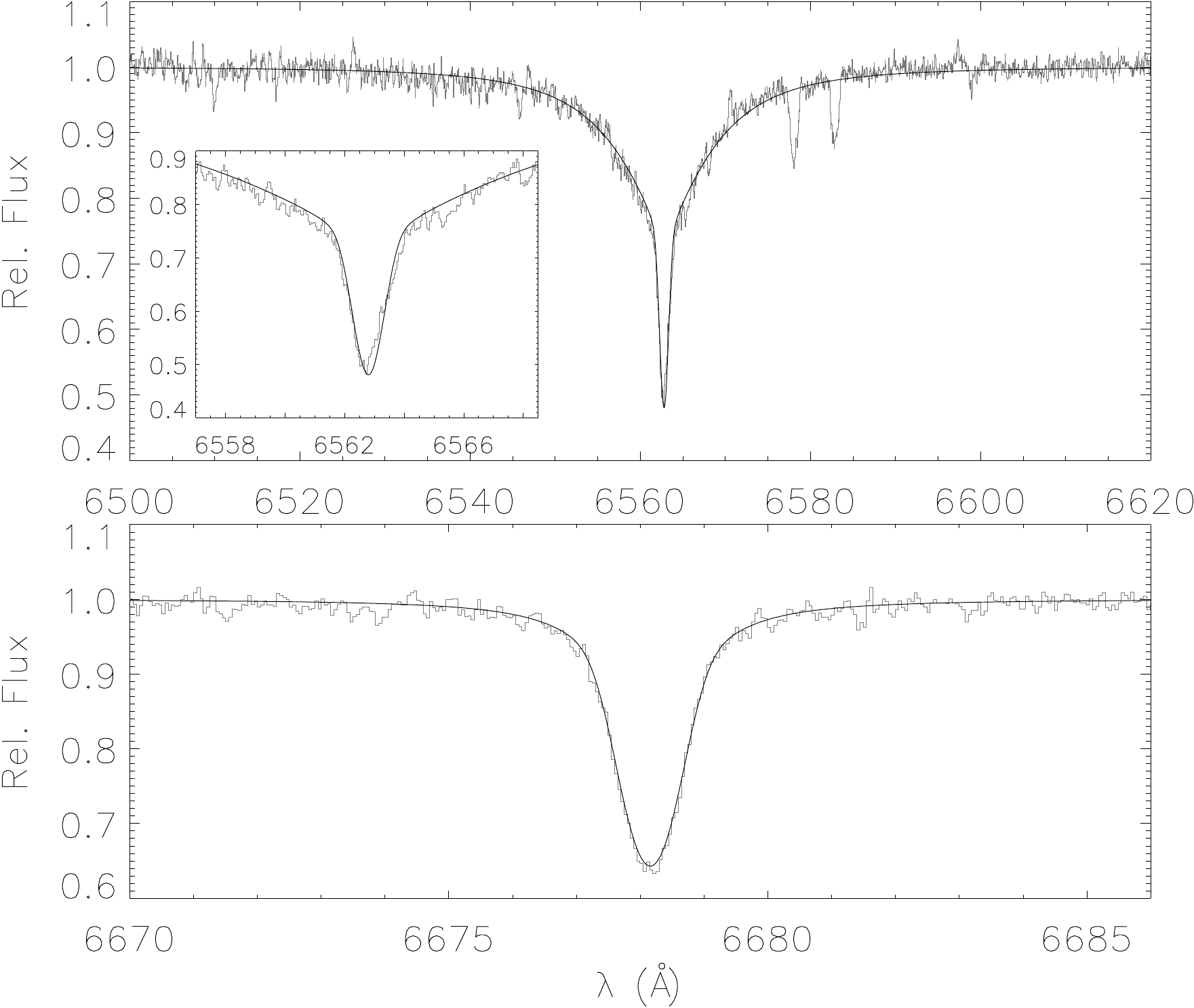}  
   \includegraphics[width=5.94cm]{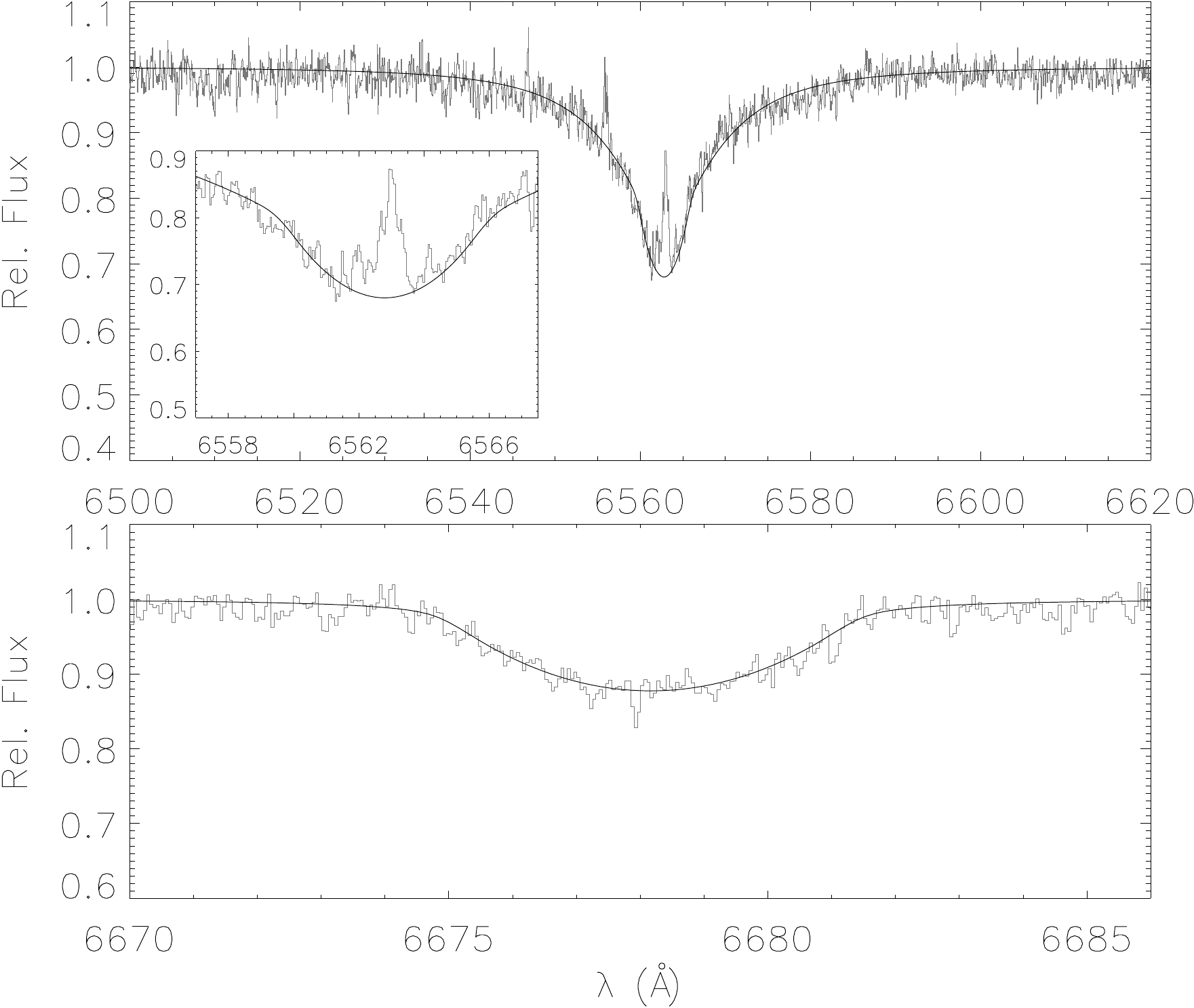} 
   \includegraphics[width=5.94cm]{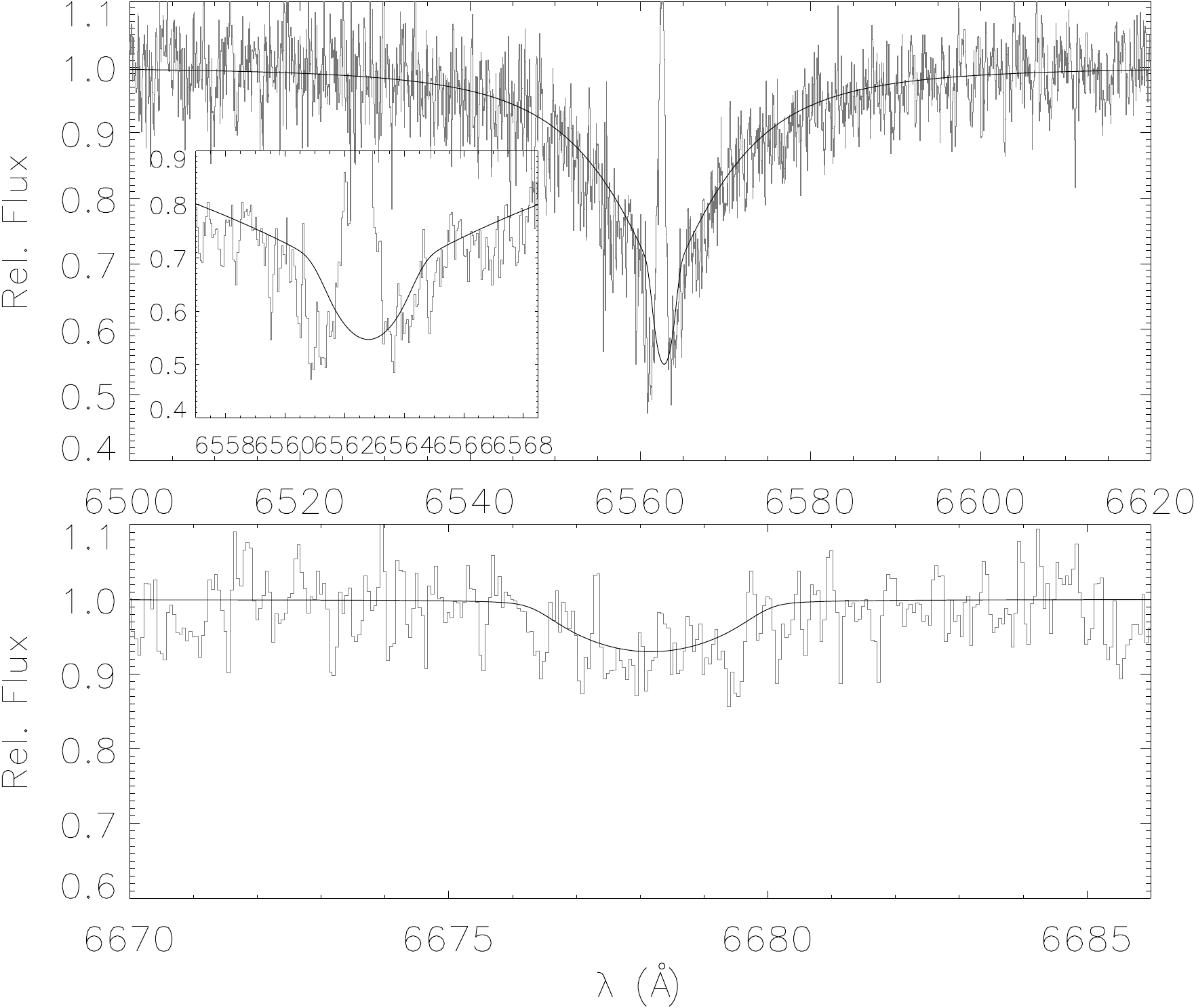} 
      \caption{
Comparison of synthetic NLTE spectra with observed spectra around the
H$\alpha$ and the He~{\sc i} line for stars at different
luminosities. We show spectra for the eMSTO stars \#118 and \#89 (left
panels), the bMS stars \#88 and \#115 (middle panels), and the rMS stars \#141, \#147
(right  panels). 
A zoom around the H$\alpha$ core is also provided in an insets for
each star.
       }
        \label{fig:fits}
   \end{figure*}

In principle, the fits provide atmospheric parameters for the sample
stars. However, with only H$\alpha$ and He\,{\sc i} $\lambda$6678\,{\AA}
available as diagnostic lines the resulting parameter values are unreliable.
In the \teff\ range covered by the sample stars one does not
only find normal stars but also a considerable fraction of chemically-peculiar 
types like Bp-, HgMn-, He-weak and He-strong stars (e.g. Smith~1996), such that
our grid covers only a fraction of the total applicable parameter space.
We therefore refrain from stating atmospheric parameter values here.
Fortunately, this has little impact on the \vsini\ determination,
as the detailed parameter-dependent line shape (not the non-LTE
effects, see above) gets washed out by the
convolution with a rotational broadening profile already above \vsini\ 
values of $\sim$50\,\kmsec\ for the S/N values achieved in the
observations. 
Our \vsini\ results for the sample stars are shown in
Table~1\footnote{Note that we found non-zero macroturbulent
  velocities for about a third of the low-\vsini\ stars (mostly bMS), following the
procedure as discussed by Firnstein \& Przybilla (2012). The
macroturbulent velocities range from about 30 to 60\,\kmsec,
comparable to the \vsini\ values themselves.}.
The listed uncertainties are again influenced by the same three factors as
stated for the radial velocity determination in
Sect.~\ref{sec:spec_data}. 
Effects of reasonable atmospheric parameter uncertainties should be covered 
within the error bars as well. Figure~\ref{fig:fits} shows examples of the
quality of the line-profile fits for the different object classes at
high and low S/N for the \vsini\ values given in Table~1.
Note that additional uncertainties to those listed in Table~1 are
introduced by continuum 
setting, which are very small in the case of the slowest rotators and
brightest targets, of the order of 1-2~\kmsec, but can be as high as
20~\kmsec\ in the faintest and in the fastest rotating stars. 

In addition to this, a number of effects could introduce
  systematic errors to the \vsini\ determination. Stellar variability
  due to e.g. low-order 
(non-radial) pulsations or spots could impact the line profiles.
Pulsational periods for Slowly Pulsating B-stars (as potentially
relevant in the spectral range investigated here) and rotational
periods of our sample stars are comparable to the duration of
our observations (three nights). The coaddition of the individual exposures
could introduce some small-scale unnoticed extra broadening of the
spectral lines, mimicking higher \vsini. While the effects are
hard to quantify, they would imply that our \vsini\ values are
slightly overestimated by tendency. Another effect comes from
undetected binarity. This will not be relevant for the bMS stars, as
their position in the colour-magnitude diagram renders relevant second
light from a companion unlikely, but the other sample stars may be
affected. The additional continuum from a (fainter) companion
will weaken the spectral lines of the primary, reducing their
depth (and not necessary widths of the spectral lines of the
  primary). 
If relevant, this would result in an underestimation of
\vsini. Overall, the effects have the potential to add some additional
systematic error to the uncertainties stated in Table~1 and shift the
values by some small amount, but they will not change the overall
rotational velocity distribution found here, and its interpretation.

A bit of caution has to be taken in the context of the
\vsini\ values for the Be stars, which, due to the complex
  profiles of their H$\alpha$, were derived from the He\,{\sc i}
  $\lambda$6678\,{\AA} line only.  
To warn the reader of the much larger uncertainties than the formal
ones listed in Table~1, \vsini\ values for Be stars have been listed
in brackets.
They were derived neglecting 
gravitational darkening and disk contamination, which however seem
less important from the recent comprehensive investigation of Ahmed \& Sigut
(2017) than implied previously (e.g. Townsend et al.~2004; Fremat et al.~2005). 
Note that the rMS stars may also be affected by gravitational
darkening to some degree, depending on how close to critical
rotational velocity  
-- in the range $\sim$400 to 500\,\kmsec\ for the sample stars --
they truely are. 

Because of all the issues discussed above (e.g.\,He gravitational
settling for the 
slow-rotators and gravitational darkening for faster rotators),
although the He line is well-detectable in almost
all the targets, we did not attempt any estimate of the He chemical
abundance. 
In the following, we will discuss separately results for MS, MSTO, Be
and binaries. 

\subsection{Main Sequence}\label{sec:MS}

\begin{figure*}
   \centering
   \includegraphics[width=18cm]{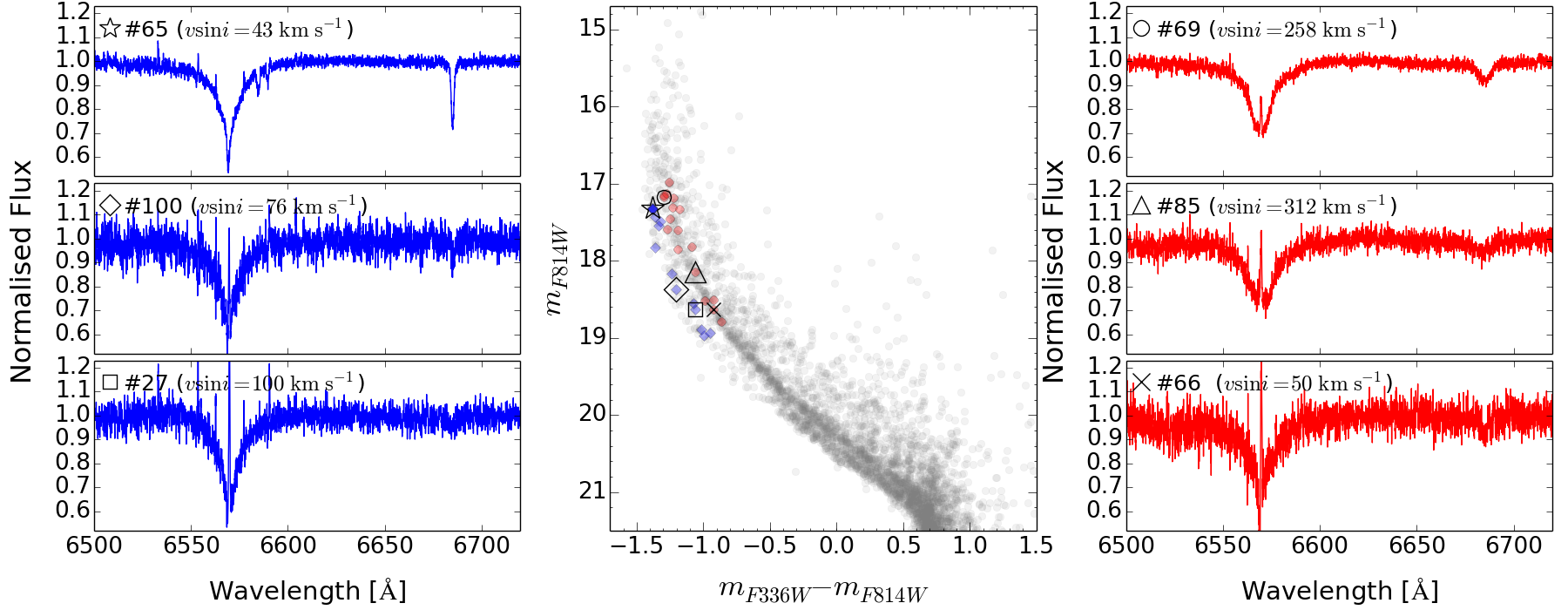}
      \caption{
  Some examples of the spectra we gathered for blue-MS (left panels)
  and red-MS stars (right panels). The represented spectra cover the whole
  range in magnitude observed for the blue-MS and red-MS stars.
  The position of all the observed MS targets on the $m_{\rm
    F814W}$-$(m_{\rm F336W}-m_{\rm F814W})$ CMD is plotted in the
  middle panel, where we explicitely indicate the location of the stars with the
  represented spectra on the side-panels.
       }
        \label{fig:rMS_bMS}
   \end{figure*}

In this section we discuss the stars on the MS of NGC\,1818,
which include the bMS and the rMS stars.
Figure~\ref{fig:rMS_bMS} displays the location on the CMD of the MS
stars with available spectroscopy. 
Our target stars are clearly distributed on the split MS of NGC\,1818,
and the association with the bMS and rMS is straightforward. 
On the side-panels we show some examples of spectra for the bMS (left)
and rMS (right) stars at different magnitudes, from the brightest to
the faintest targets, to give an idea of the quality of our data. 
The spectra show both the H$\alpha$ and the He~{\sc i} line. 
Note that we do not provide any \vsini\ for the bMS star \#107, which
displays strong emission in wings. This star does not have a peculiar
position on the CMD, being located on the bMS.
The two rMS stars \#102 and \#26 display some core emission.
  
A qualitative inspection of this figure immediately suggests a different
projected rotation for the two brightest stars, at $m_{\rm F814W}
\sim$17.2~mag. Stars \#65 and \#69 have similar luminosity, but a
different color, and, as expected, the spectrum of rMS star \#69 is
consistent with a significanly higher rotation than the bMS \#65, as
indicated by the much broadened profiles of both the H$\alpha$ and the
He~{\sc i} line. 
Similarly the spectral profiles of star \#85 are broader than those of
the bMS star \#100, which has a similar magnitude. 
Lower luminosity stars invert the trend, with the rMS \#66 having
lower \vsini\ than the bMS \#27. 
Keeping in mind that the \vsini\ values are lower limits to the real
rotations of the star, depending on the inclination $i$, it is more
likely that a fast-rotating star has a higher \vsini\ than a slow
rotator, but there is some possibility to have the opposite.
However, note that larger uncertainties are associated with the
lower-luminosity stars with poorer S/N.

The mean \vsini\ we obtain for the 14 bMS and the 17 rMS are 
$<$\vsini$>_{\rm bMS}$=71$\pm$10~\kmsec\ ($\sigma$=37~\kmsec) and
$<$\vsini$>_{\rm rMS}$=202$\pm$23~\kmsec\ ($\sigma$=91~\kmsec),
respectively. This means that our data are consistent with a
difference in the rotation regimes of the two MSs of NGC\,1818, with
an average difference $<\Delta_{\rm
  {\vsini_{rMS}-\vsini_{bMS}}}>$=$+$131$\pm$25~\kmsec, a $>$5~$\sigma$
difference.  

\begin{figure*}
   \centering
   \includegraphics[width=17.5cm]{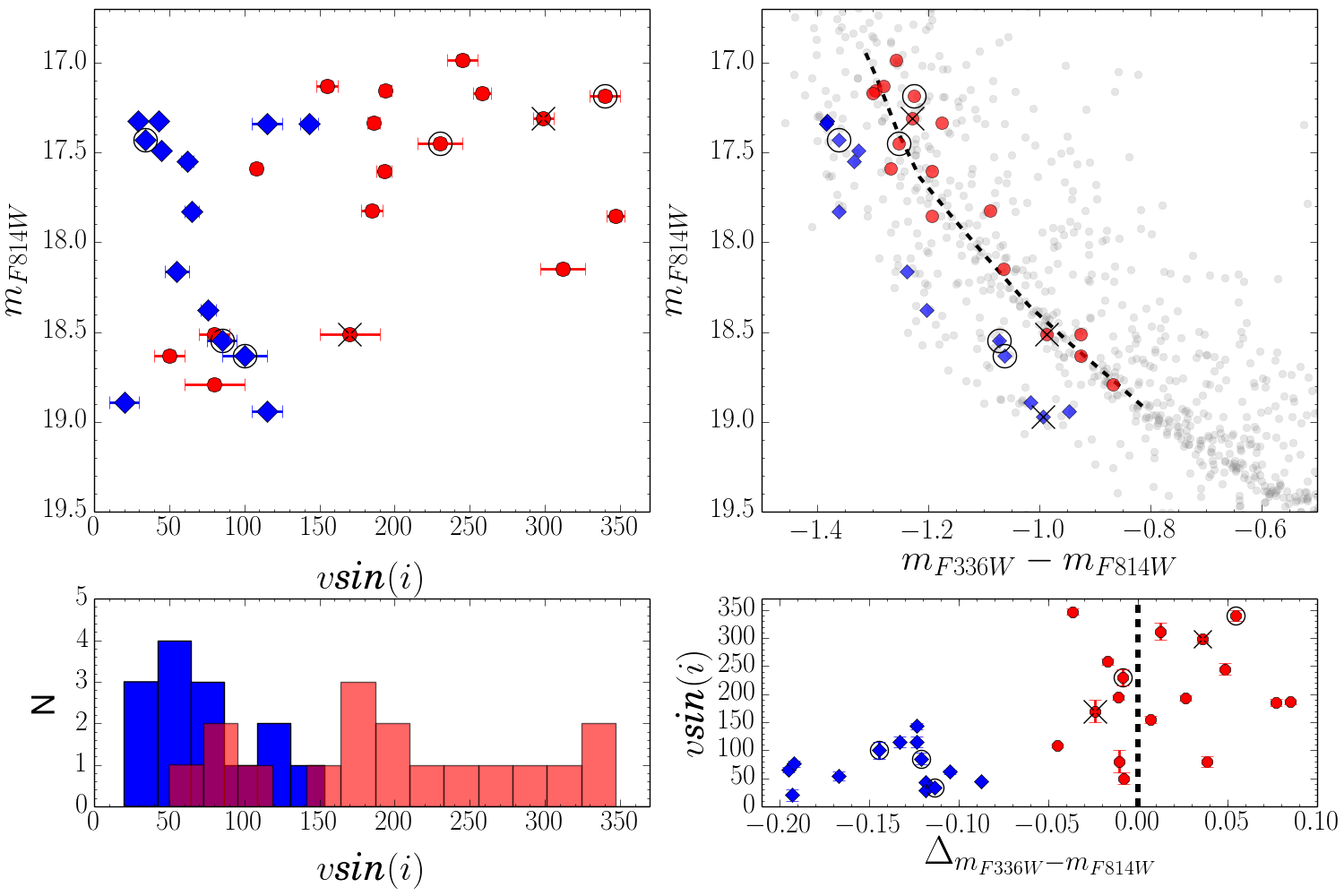}
      \caption{
{\it Left panels}: $m_{\rm F814W}$ mag as a function of the projected rotational
velocities \vsini\ for the analysed MS stars in NGC\,1818 (upper panel); 
on the lower panel we show the histogram distribution of \vsini\ for
rMS (red) and bMS (blue).
{\it Right panels}: $m_{\rm F814W}$-$(m_{\rm F336W}-m_{\rm F814W})$
CMD of NGC\,1818 zoomed on the split MS. The black-dashed line is a
fiducial for the rMS (upper panel). 
The lower panel shows \vsini\ as a function of the difference between
the $(m_{\rm F336W}-m_{\rm F814W})$ colour of each MS star and the
colour of the fiducial ($\Delta_{m_{\rm F336W}-m_{\rm F814W}}$). The
black-dashed line is a  $\Delta_{m_{\rm F336W}-m_{\rm F814W}}$=0.
In all the panels rMS and bMS have been represented with red-filled
circles, and blue diamonds, respectively. Circled and
crossed symbols indicate SB1 candidates and stars with H$\alpha$ core and/or
wings emissions, respectively. 
       }
        \label{fig:rot_MS}
   \end{figure*}

Our results on MS stars are outlined in Fig.~\ref{fig:rot_MS}. 
The histogram distribution of \vsini\ for rMS and bMS stars, shown
in the lower-left panel, illustrates the much wider range in \vsini\
for the rMS stars. 
The left-upper panel displays the $m_{\rm F814W}$ as a function of \vsini\
for rMS and bMS stars.
In all the covered range in $m_{\rm F814W}$, bMS stars are shown to be slower
rotators, but are consistent with a non-zero rotation in all cases,
with some hints of higher rotation for fainter magnitudes. 
On the other hand, the spread in \vsini\ for the rMS is higher, and
stars fainter than $m_{\rm F814W} \sim$18.5~mag seem to rotate slower,
with values close to the bMS stars. We suggest caution however in the
interpretation of the these fainter stars, whose errors in \vsini\ are
much larger. 

In the right panels of Fig.~\ref{fig:rot_MS} we analyse the colour of
the MS stars as a function of the \vsini\ values.
To this aim we have drawn a fiducial line by eye defining the rMS on
the $m_{\rm F814W}$-$(m_{\rm F336W}-m_{\rm F814W})$ CMD (upper-right panel).
The difference in colour between each analysed MS star and the
fiducial, $\Delta_{m_{\rm F336W}-m_{\rm F814W}}$, does not show any
significant trend with \vsini\ (lower-right panel).  

\subsection{Extended MSTO}\label{sec:MSTO}

\begin{figure*}
   \centering
   \includegraphics[width=18cm]{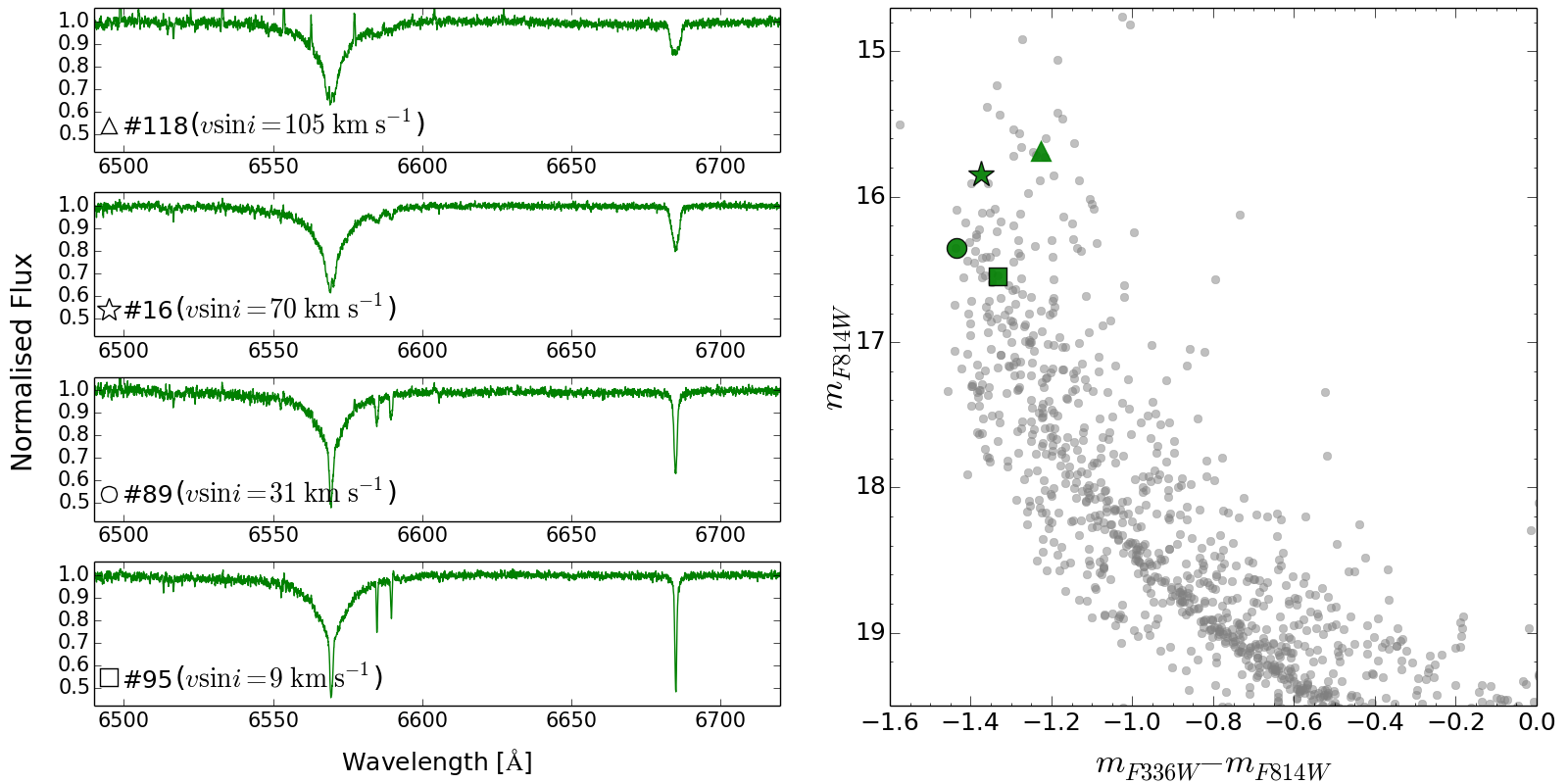}
      \caption{
{\it Left panels}: Spectra of the stars classified as MSTO. For each
star we report the inferred \vsini\ value.
{\it Right panel}: $m_{\rm F814W}$-$(m_{\rm F336W}-m_{\rm F814W})$ CMD
of NGC\,1818 zoomed around the MSTO region. The location of the
spectroscopically-observed stars is displayed in green symbols.
Each symbol indicates the star with the corresponding symbols reported
on the left-side spectra.
       }
        \label{fig:MSTO_CMD}
   \end{figure*}

\begin{figure*}
   \centering
   \includegraphics[width=14cm]{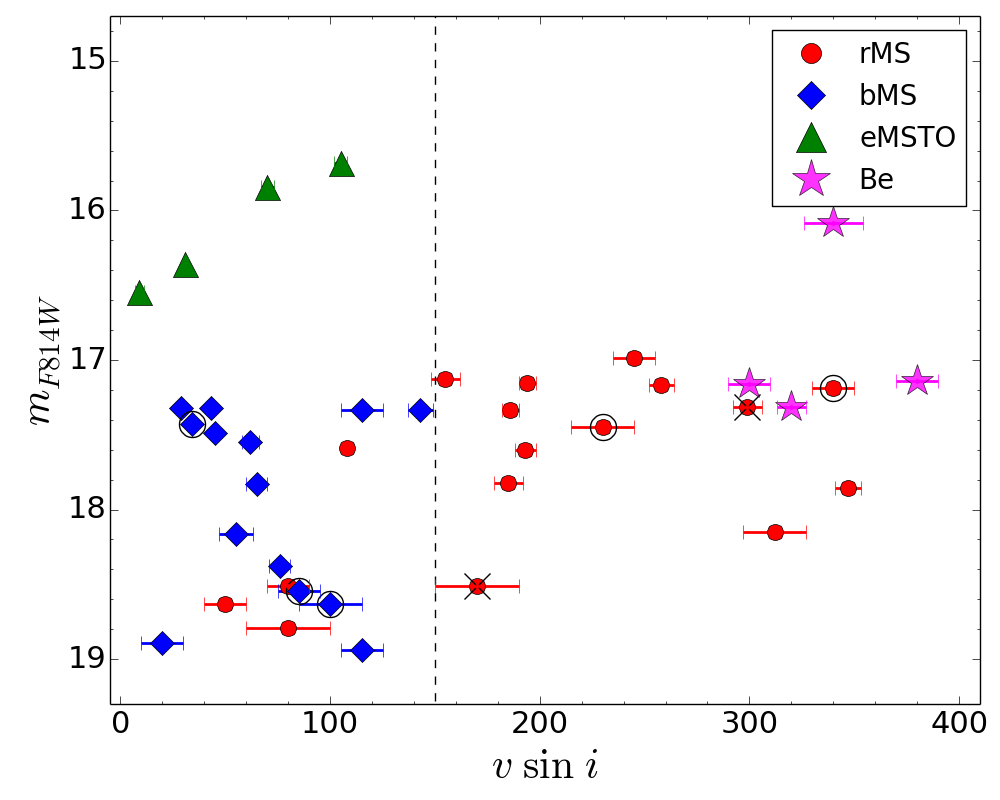}
      \caption{
Maginitude $m_{\rm F814W}$ as a function of \vsini\ [\kmsec] for rMS
(red circles), bMS (blue diamonds), eMSTO stars (green triangles), and
Be stars (magenta stars).
       }
        \label{fig:MS_MSTO}
   \end{figure*}

Figure~\ref{fig:MSTO_CMD} displays the position on the CMD of the four
eMSTO stars, and their spectra.
The distribution of the inferred \vsini\ values for these stars,
compared to the bMS and rMS ones, is shown in Fig.\ref{fig:MS_MSTO}.
At least for the objects studied here, none of the eMSTO stars reaches
\vsini\ values higher than $\sim$150\kmsec, which is the limiting
value above which we do not find any bMS star.
Of course, this does not necessarily mean that none of the analysed eMSTO is the
counterpart of rMS stars, as below  $\sim$150\kmsec\ we observe both
rMS and bMS.

\subsection{Be Stars}\label{sec:Be}

\begin{figure*}
   \centering
   \includegraphics[width=17.5cm]{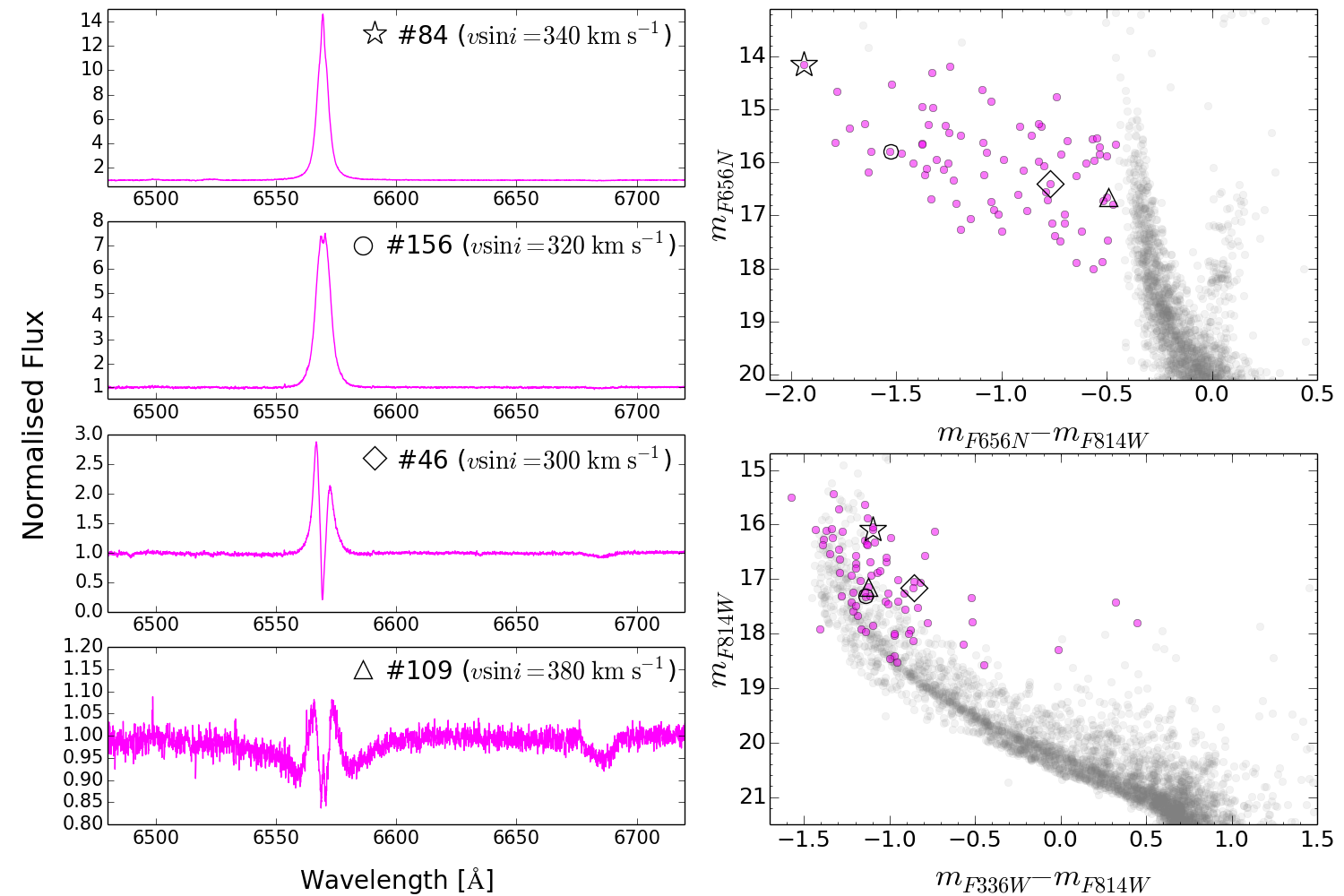}
      \caption{
       Spectra of the four Be stars in NGC\,1818 (left panel), and their
       location plotted with different black open symbols on the 
       $m_{\rm F656N}$-$(m_{\rm F656N}-m_{\rm F814W})$ (top-right
       panel) and $m_{\rm F814W}$-$(m_{\rm F336W}-m_{\rm F814W})$
       (bottom-right panel) CMDs.
       From the top to the bottom the spectra represent stars with higher
       excess in $m_{\rm F656N}$, corresponding to a more pronounced
       emission in H$\alpha$.
       In both CMDs all the selected Be stars have been plotted as
       magenta dots.
       }
        \label{fig:Be}
   \end{figure*}

Figure~\ref{fig:Be} displays the spectra of the four stars selected as
Be stars. The top-right panel illustrates the position of all the Be candidates on
the $m_{\rm F656N}$-$(m_{\rm F656N}-m_{\rm F814W})$ CMD, e.g. Be stars are those
with an excess in $m_{\rm F656N}$ due to the emission in the H$\alpha$. 
We have selected the four Be-candidates based on their location on the
$m_{\rm F656N}$-$(m_{\rm F656N}-m_{\rm F814W})$ CMD\footnote{As noted
  by Dupree et al.\,(2017), relying on the $HST$ photometry using the narrow filter
$m_{\rm F656N}$ to identify the Be stars might result in missing of
some of them, given the high radial velocities of our stars.}, hence 
our spectra in turn confirm that the stars identified as Be on our CMD
have H$\alpha$ emission.
The right-bottom panel of Fig.~\ref{fig:Be} shows that on the $m_{\rm
    F814W}$-$(m_{\rm F336W}-m_{\rm F814W})$ CMD Be stars are spread in
  a relatively wide range in color at $m_{\rm F814W}$ brighter than
  $\sim$18.5. 
As noted in Milone et al.\,(2018) nearly all of them are either located
on the rMS or populate the reddest part of the eMSTO.

As in Rivinius et al.\,(2013) Be stars are {\it ``very rapidly rotating
main sequence B stars, which, through a still unknown, but
increasingly constrained process, form an outwardly diffusing gaseous,
dust-free Keplerian disk.''}
Hence, the shape of the H$\alpha$ profile gives us information on the
inclination $i$. 
Star \#84, which is the brightest in $m_{\rm F656N}$, is the one
which appears closer to pole-on orientation, as it has a strong narrow
emission, and no absorption. Star \#156 also shows a H$\alpha$
morphology consistent with a near pole-on view, though with a reduced
emission. The double-peaked H$\alpha$ profiles with central absorption
of the other Be stars \#46 and \#109 imply a higher $i$, i.e. we see
these objects through their disk (edge-on).

As discussed in Sect.~\ref{sec:rotation}, the
  \vsini-determination for the Be stars has to rely solely on
He\,{\sc i} 6678\,{\AA}. The values inferred are the highest
in our sample, being all $\gtrsim$300\,\kmsec\ (see Fig.~7).
While it is expected that Be stars are fast rotators, the more pole-on
view of \#86 and \#156 would suggest that they will likely rotate
very close to their critical velocity. We also note that
star \#46 has an asymmetric He\,{\sc i} profile, which either could be
a hint for a second light component (the secondary in a physical
binary, or a chance alignment) or the signature of inhomogeneities in
the disk.

\subsection{Binaries}\label{sec:binaries}

We did not attempt any \vsini\ determination for stars displayig a
double-lined spectrum. These stars could be either physical binaries
or aligments of stars.
However, it could be interesting to investigate the position of these stars
on the CMD. 

\begin{figure*}
   \centering
\includegraphics[width=17.5cm]{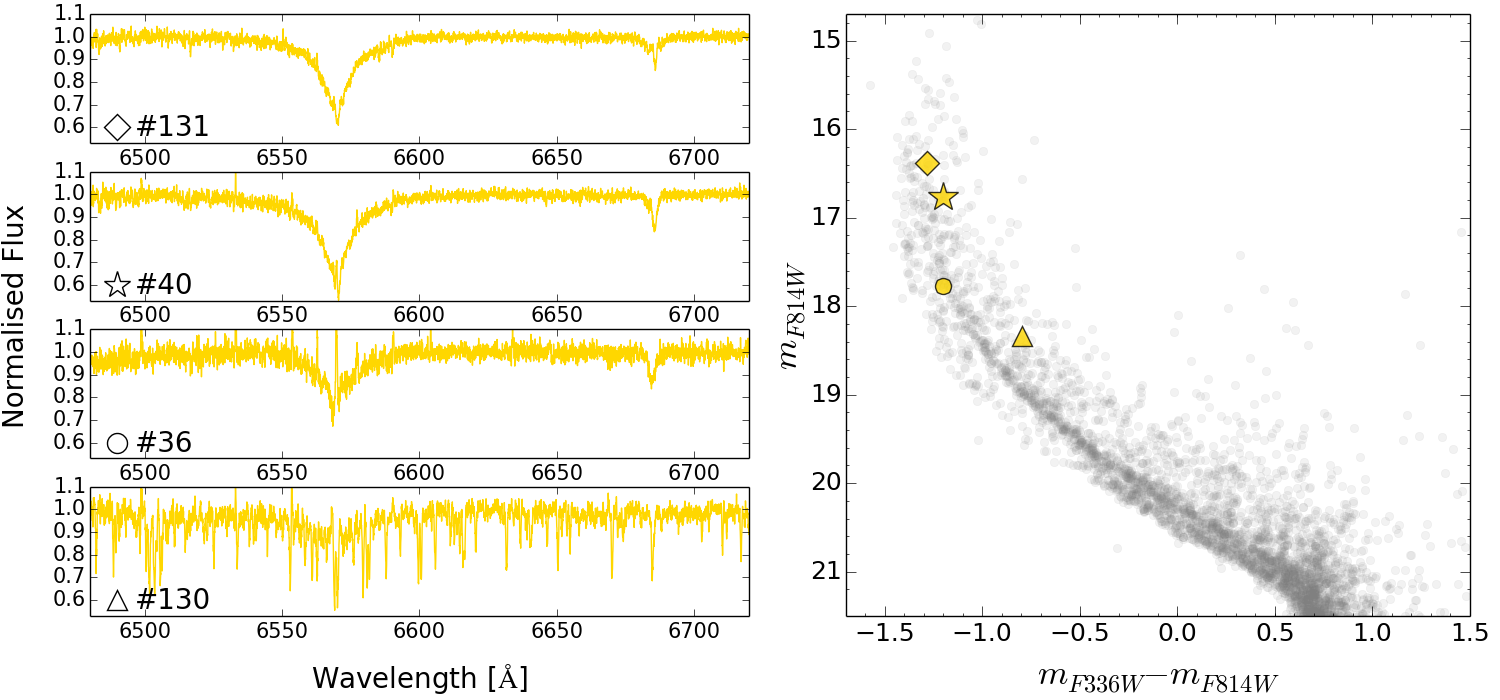}
      \caption{
Spectra (left) and position on the $m_{\rm F814W}$-$(m_{\rm
  F336W}-m_{\rm F814W})$ CMD of the stars classified as binaries from
their double-lined spectra. The different yellow symbols on the CMD
correspond to different spectra on the left side. 
       }
        \label{fig:binaries}
   \end{figure*}

Figure~\ref{fig:binaries} shows the location of the double-lined stars
on the $m_{\rm F814W}$-$(m_{\rm F336W}-m_{\rm F814W})$ CMD of NGC\,1818.
Two of them, \#40 and \#130, occupy indeed the sequence of binaries,
redder than the rMS. On the other hand, star \#36 is located on the
rMS: as discussed above we expect some contamination between the rMS
and the binary region. The more luminous star \#131 could be
classified as a MSTO.

\section{Summary and Discussion}\label{sec:discussion}

We have derived projected rotational velocities \vsini\ for rMS and
bMS, eMSTO, and Be stars in the young cluster NGC\,1818.
The presented analysis suggests:
\begin{itemize}
\item{a clear difference in the mean \vsini\ between bMS and rMS
    stars, having average values of $<$\vsini$>_{\rm
      bMS}$=71$\pm$10~\kmsec\ ($\sigma$=37~\kmsec) and
    $<$\vsini$>_{\rm rMS}$=202$\pm$23~\kmsec\ ($\sigma$=91~\kmsec),
    respectively;}
\item{a lower average \vsini\ ($<$\vsini$>$=54$\pm$25~\kmsec)
    for the four stars classified as eMSTO, more
    consistent with the values measured for the bMS stars;}
\item{high \vsini\ (average $<$\vsini$>$=335$\pm$20~\kmsec), the
    highest in our sample, for the analysed Be stars.} 
\end{itemize}

Our results provide the first direct evidence from spectroscopy of
different rotation rates for the two MSs observed in a young cluster.

Although \vsini\ values are lower limits to the real surface rotational
velocity of the stars, they allow us to draw some constraints to the
multiple stellar populations observed in young star clusters,
specificaly in NGC\,1818.  
The different mean \vsini\ derived for bMS and rMS stars confirms the
suggestion coming from the analysis of high-precision 
CMDs from $HST$ that the split MS of young GCs is connected with
different rotational regimes.
Indeed, Milone et al.\,(2018) have shown that the CMD of NGC\,1818 is
best-fitted by isochrones with rotation $\omega$=0 (representative of
models with low rotation velocity) to reproduce the bMS, and with
$\omega$=0.9~$\omega_{\rm crit}$ (representative of high initial
rotation) to fit the rMS (see their Fig. 10 for the isochrone fit to
the CMD).  

The Geneva models for $\omega$=0.9~$\omega_{\rm crit}$ corresponding
to the masses evolving in this 
cluster ($\sim$8$M_{\odot}$) display surface rotation
velocities of 330-340\kmsec\ during the MS lifetime, consistent with the maximum
\vsini\ in our sample, 347$\pm$6 measured for the rMS star \#141, in
the hypothesis that this star has inclination close to $90^{\circ}$\footnote{
Actually these surface velocities correspond to a ratio $\omega \sim$0.8~$\omega_{\rm crit}$. Such a difference between the formal value given in the database and the actual value in the models is due to the passage from solid-body rotation assumed in the initial models before the ZAMS to differential rotation on the MS. An equilibrium profile of $\omega$ is built inside the star at the expense of the surface rotation velocity (Ekstr\"{o}m, private communication).}.
The lower values of \vsini\ that we derived for the other stars are
consistent with velocities of $\sim$330-340~\kmsec\ and inclination  $<90^{\circ}$.

By simply scaling down the maximum \vsini\ observed for our MS stars
and assuming that it corresponds to $\omega$=0.9~$\omega_{\rm crit}$,
the average \vsini\ of the bMS stars (71~\kmsec) would correspond to a 
value of $\omega \sim$0.2~$\omega_{\rm crit}$. Although higher than zero,
this is indeed a quite low velocity, still compatible with the
location of the blue main sequence, if one looks at the Geneva
isochrone location. 
However, being the average \vsini\ values lower limits to the real
stellar rotation, it is likely that the bMS is populated by stars with
$\omega >$0.2~$\omega_{\rm crit}$.

It has been suggested that the bMS is composed of ``braked'' stars, and
not simply of stars born as slowly-rotators (D'Antona et al.\,2017).
In this scenario, the upper MS stars are assumed to be ``nuclearly''
younger (they define indeed a younger isochrone (Milone et al.\,2018),
because, for most of their MS lifetime, they have been subject to the
core-envelope mixing of rapidly rotating stars. 

Our bMS \vsini\ values show a possible increase of rotation towards
fainter magnitudes, which, if real, would be consistent with the
braking hypothesis, because, approximately above the TO of the
non-rotating isochrone, we may find in MS only stars almost fully
braked, otherwise they would have been already evolved. The same
reasoning also applies to our four eMSTO stars: none shows \vsini\ as
high as the rMS, possibly suggesting a lack of fast rotators below the
highest luminosity turnoff. Although based on a small sample size
(four stars), a hint of lower velocity for dimmer magnitudes is
suggested from our \vsini\ of eMSTO stars. This is expected if the
eMSTO is interpreted in terms both of braking and of different
rotation rates, but note that the post-TO evolution itself implies a
braking of the envelope. 

Clearly, the analysed Be stars are the fastest-rotators in our sample, with
\vsini\ overall higher than those observed along the rMS. 
In all of them we detect strong H$\alpha$ emission, clearly indicating
a fast-rotating star with a Keplerian decretion disk (Rivinius et
al.\,2013). The generally higher \vsini\ values for the Be stars suggest that
these objects are the closest to the breaking speed rotation in the cluster.  
An increase of the ratio $\omega/\omega_{\rm crit}$ occurs during the MS lifetime in
rotational models, although the equatorial velocity is reduced, and
this may be the reason of the formation of the decretion disk.

Qualitatively, our work strongly supports a deep influence of stellar
rotation on the appearence of multiple and/or spread photometric
sequences and on shaping the CMD of young stellar clusters. In more
details, our measurements agree with a rotation for the rMS of about
$\omega$=0.8~$\omega_{\rm crit}$, which, in fact, is the rotation
rate in the detailed Geneva models formally labelled as
$\omega$=0.9~$\omega_{\rm crit}$ (see footnote 4).  
A rotation much closer to the critical value might be attributed to the Be stars.
We also find a non-zero rotation for bMS
stars, which rotate at $\omega \gtrsim$0.2~$\omega_{\rm crit}$, 
anyway much slowly than rMS stars.
We also note that the larger spread in the \vsini\ values for rMS
stars, with respect to the bMS,
might only be partially ascribed to the different inclinations {\it i}
coupled with high intrinsic rotations, and we cannot exclude some spread
in the stellar rotation for these stars.

The presence of split MS in young clusters is one of the
most-intriguing and unexpected discoveries of the last years in the
field of stellar populations.
  This work needs to be seen as the first exploration of some
phenomena that could affect not only the cluster formation but also
stellar-evolution modeling.
Further studies of the rotation velocities in additional clusters,
with equivalent or larger samples, will be crucial to fully understand
what is clearly a complex phenomenon.   

\begin{deluxetable}{ccccccccc}
\tablewidth{0pt}
\tablecaption{Observation details, coordinates and radial velocities for our spectroscopic targets.\label{tab:data}}
\tablehead{
  ID       & $\alpha$(2000)& $\delta$(2000)  & $m_{F336W}$ & $m_{F656N}$ & $m_{F814W}$ &  RV$_{\rm helio}$ [\kmsec] & \vsini\ [\kmsec]& CMD~region    
}
\startdata
 31 &  05:04:03.72 & $-$66:25:30.8 & 16.3236 & 17.2822 & 17.591 & 311$\pm$2 & 108$\pm$2   & redMS   \\	 
 33 &  05:04:14.16 & $-$66:25:34.0 & 15.8604 & 16.8372 & 17.155 & 314$\pm$3 & 194$\pm$4   & redMS   \\     
 83 &  05:04:17.44 & $-$66:26:17.0 & 15.8471 & 16.7573 & 17.127 & 310$\pm$2 & 155$\pm$7   & redMS   \\     
117 &  05:04:18.22 & $-$66:26:40.2 & 16.4095 & 17.3158 & 17.603 & 307$\pm$2 & 193$\pm$5   & redMS   \\     
102 &  05:04:10.48 & $-$66:26:24.7 & 16.0832 & 16.9004 & 17.311 & 310$\pm$5 & 299$\pm$7   & redMS   \\     
141 &  05:04:00.86 & $-$66:26:51.2 & 16.6612 & 17.4815 & 17.855 & 309$\pm$5 & 347$\pm$6   & redMS   \\     
 69 &  05:04:07.76 & $-$66:26:01.0 & 15.8713 & 16.8311 & 17.170 & 310$\pm$2 & 258$\pm$6   & redMS   \\     
 85 &  05:04:01.03 & $-$66:26:13.3 & 17.0840 & 17.7970 & 18.148 & 314$\pm$8 & 312$\pm$15  & redMS   \\     
 26 &  05:04:14.83 & $-$66:25:24.0 & 17.5280 & 18.2406 & 18.514 & 315$\pm$12& 170$\pm$20  & redMS   \\     
 66 &  05:04:20.57 & $-$66:26:03.2 & 17.7081 & 18.3423 & 18.634 & 315$\pm$8 &  50$\pm$10  & redMS   \\     
 57 &  05:04:07.91 & $-$66:25:51.8 & 16.1990 & 17.1258 & 17.451 & 297$\pm$6 & 230$\pm$15  & redMS   \\     
 15 &  05:04:08.98 & $-$66:25:16.7 & 17.9261 & 18.4930 & 18.794 & 313$\pm$10&  80$\pm$20  & redMS   \\     
147 &  05:04:03.92 & $-$66:26:58.5 & 17.5850 & 18.1842 & 18.510 & 315$\pm$10&   80$\pm$10 & redMS   \\     
 77 &  05:04:05.48 & $-$66:26:07.0 & 15.9587 & 16.8638 & 17.184 & 296$\pm$5 &  340$\pm$10 & redMS  \\
133 &  05:04:18.22 & $-$66:26:49.9 & 16.1560 & 17.0691 & 17.332 & 305$\pm$3 &  186$\pm$4  & redMS  \\
157 &  05:04:10.33 & $-$66:27:16.0 & 16.7341 & 17.4968 & 17.823 & 314$\pm$6 &  185$\pm$7  & redMS  \\
 27 &  05:04:13.42 & $-$66:25:20.9 & 17.5688 & 18.3155 & 18.632 & 323$\pm$8 & 100$\pm$15  & blueMS  \\
 63 &  05:04:16.57 & $-$66:25:47.6 & 16.4660 & 17.4731 & 17.826 & 315$\pm$2 &  65$\pm$5   & blueMS  \\
 65 &  05:04:17.58 & $-$66:26:00.1 & 15.9411 & 16.9562 & 17.322 & 307$\pm$2 &  43$\pm$2   & blueMS  \\
107 &  05:04:23.40 & $-$66:26:30.2 & 17.9798 & 18.7799 & 18.973 &   --      & --          & blueMS  \\
115 &  05:04:14.26 & $-$66:26:36.3 & 15.9530 & 17.0013 & 17.337 & 313$\pm$3 & 143$\pm$6   & blueMS  \\
148 &  05:04:10.04 & $-$66:26:59.9 & 16.1658 & 17.1569 & 17.491 & 310$\pm$2 &  45$\pm$2   & blueMS  \\
 56 &  05:04:04.97 & $-$66:25:48.8 & 17.9942 & 18.6954 & 18.941 & 315$\pm$12& 115$\pm$10  & blueMS  \\
 28 &  05:04:18.09 & $-$66:25:30.5 & 17.4755 & 18.3383 & 18.547 & 290$\pm$6 &  85$\pm$10  & blueMS  \\
 88 &  05:04:17.58 & $-$66:26:00.1 & 15.9411 & 16.9562 & 17.322 & 304$\pm$2 &  29$\pm$2   & blueMS  \\
 96 &  05:04:16.91 & $-$66:26:24.6 & 16.0693 & 17.1000 & 17.430 & 288$\pm$1 &  34$\pm$2   & blueMS  \\
166 &  05:04:14.26 & $-$66:26:36.3 & 15.9530 & 17.0013 & 17.337 & 311$\pm$2 & 115$\pm$10  & blueMS  \\
 20 &  05:04:06.97 & $-$66:25:20.5 & 16.2154 & 17.2305 & 17.549 & 312$\pm$2 &  62$\pm$4   & blueMS  \\
140 &  05:04:21.07 & $-$66:26:52.1 & 17.8749 & 18.6517 & 18.890 & 307$\pm$10&  20$\pm$10  & blueMS  \\
152 &  05:04:17.47 & $-$66:27:07.2 & 16.9257 & 17.8575 & 18.164 & 320$\pm$5 &  55$\pm$8   & blueMS  \\
100 &  05:04:03.24 & $-$66:26:21.5 & 17.1726 & 18.0690 & 18.375 & 312$\pm$6 &  76$\pm$5   & blueMS  \\
 41 &  05:04:12.64 & $-$66:25:40.2 & 15.7297 & 16.6497 & 16.987 & 306$\pm$2 & 245$\pm$10  & MSTO    \\     
 95 &  05:04:06.24 & $-$66:26:20.6 & 15.2109 & 16.2206 & 16.544 & 310$\pm$1 &   9$\pm$2   & MSTO    \\     
 16 &  05:04:12.99 & $-$66:25:15.7 & 14.4708 & 15.4789 & 15.846 & 306$\pm$1 &  70$\pm$3   & MSTO    \\     
 89 &  05:04:08.61 & $-$66:26:16.2 & 14.9200 & 16.0089 & 16.355 & 307$\pm$1 &  31$\pm$1   & MSTO    \\     
118 &  05:04:21.49 & $-$66:26:39.9 & 14.4558 & 15.2953 & 15.683 & 308$\pm$1 &  105$\pm$3  & MSTO    \\     
 46 &  05:04:08.96 & $-$66:25:41.2 & 16.3038 & 16.3930 & 17.162 & 309$\pm$5 & [300$\pm$10]& Be      \\
156 &  05:04:15.29 & $-$66:27:12.9 & 16.1724 & 15.7900 & 17.315 & 318$\pm$7 & [320$\pm$7] & Be      \\
109 &  05:04:11.97 & $-$66:26:34.0 & 16.0199 & 16.6490 & 17.142 & 314$\pm$8 & [380$\pm$10]& Be      \\
 84 &  05:04:20.27 & $-$66:26:14.8 & 14.9870 & 14.1445 & 16.085 & 292$\pm$5 & [340$\pm$14]& Be      \\
131 &  05:04:11.87 & $-$66:26:44.4 & 15.0971 & 16.0634 & 16.381 &     --    &   --        & Binary candidate    \\     
 40 &  05:04:07.32 & $-$66:25:38.2 & 15.5616 & 16.4533 & 16.762 &     --    &   --        & Binary candidate    \\     
130 &  05:04:07.00 & $-$66:26:44.4 & 17.5428 & 18.1193 & 18.339 &     --    &   --        & Binary candidate \\
 36 &  05:04:19.68 & $-$66:25:39.2 & 16.5695 & 17.4825 & 17.769 &     --    &   --        & Binary candidate   \\     
\enddata
\end{deluxetable}

%
%

%
%

%
%

%
%

%
%

%
%

%
%

%
%

%
%

\acknowledgments
The authors are grateful to the anonymous referee for useful discussion.
AFM and GDC acknowledge support by the Australian Research Council through
Discovery Early Career Researcher Award DE160100851 and Discovery
project DP150103294. 
APM has been supported by the European Research Council through the
Starting Grant ``GALFOR'' and the FARE-MIUR project R164RM93XW ``SEMPLICE''.

\software{{\sc Atlas9} (Kurucz 1993), {\sc IRAF} (Tody 1986, Tody
  1993), {\sc Detail}, {\sc Surface} (Giddings 1981; Butler \&
  Giddings 1985), {\sc Spas} (Hirsch 2009)}

\bibliographystyle{aa}

\end{document}